\documentclass[aps,pre,reprint,floatfix,superscriptaddress]{revtex4-1}
\usepackage{graphicx}
\usepackage[latin1]{inputenc}
\usepackage{amsmath}
\usepackage{amsfonts}
\usepackage{amssymb}
\usepackage{xcolor}

\begin{document}

\title{Formation and crossover of multiple helical dipole chains}

\date{\today}

\author{Ansgar~Siemens}
\email{asiemens@physnet.uni-hamburg.de}
\affiliation{Zentrum f\"ur Optische Quantentechnologien, Fachbereich Physik, Universit\"at Hamburg, Luruper Chaussee 149, 22761 Hamburg Germany} 
\author{Peter~Schmelcher}
\email{pschmelc@physnet.uni-hamburg.de}
\affiliation{Zentrum f\"ur Optische Quantentechnologien, Fachbereich Physik, Universit\"at Hamburg, Luruper Chaussee 149, 22761 Hamburg Germany}
\affiliation{Hamburg Center for Ultrafast Imaging, Universit\"at Hamburg, Luruper Chaussee 149, 22761 Hamburg Germany}

\begin{abstract}

\noindent We investigate the classical equilibrium properties and metamorphosis of the ground state of interacting dipoles with fixed locations on a helix.  
The dipoles are shown to align themselves along separate intertwined dipole chains forming single, double, and higher-order helical chains. 
The number of dipole chains, and their properties such as chirality and length scale on which the chains wind around each other, can be tuned by the geometrical parameters. 
We demonstrate that all possible configurations form a self-similar bifurcation diagram which can be linked to the Stern-Brocot tree and the underlying Farey sequence. 
We describe the mechanism responsible for this behavior and subsequently discuss corresponding implications and possible applications.

\end{abstract}

\maketitle

\paragraph*{Introduction}--- 
Efforts of miniaturization of functional devices have been progressing steadily in the last decades. 
Due to advances in material science, it is now possible to manufacture a plethora of one-dimensional (1D) nano-materials for experimental use \cite{rodriguez1995,lau2006,lan2011,samuelson2004,ijima1993,ivanov1994,shaikjee2012}. 
An especially intriguing class of 1D materials are chains of particles with permanent dipole moments, since it is possible to controllably encode information \cite{vedmedenko2014} and transfer energy along a linear chain \cite{deleeuw1998,sim1999,dejonge2004,dejonge2007,zampetaki2018a,fateev2002} by exciting the orientation of the dipoles from an equilibrium state. 
It has also been shown that dipole chains can be assembled into logic gates \cite{dejonge2006}, allowing to steer the energy transfer. 
Furthermore, simulations show that they could be used as waveguides that are able to transport signals below the diffraction limit \cite{kornovan2020,maier2003,brongersma2000}, thereby overcoming size limits for guiding and modulating light.

Realizing such quasi-1D molecular arrays in experimental studies is possible \cite{huang2010,jerome1988,weckesser2001}, for example through self-organization \cite{schnadt2008a} or by artificial creation through various lithography methods \cite{hersam2000}. 
Usually, the studies of molecular arrays focus on arrangements on planar surfaces or studies of bulk materials - with more sophisticated geometric configurations being muss less explored. 
From a material science point of view, these more sophisticated structures can possess several advantages, such as the increased stability and resistance to deformation observed in helical nanostructures \cite{kornyshev2007,marko1994}.
Elaborate three-dimensional setups of dipoles could for example be realized in the context of metal-organic frameworks (MOF) \cite{gonzalez-nelson2019}, which are materials consisting of inorganic building units (metal ions) that are linked with organic molecules. 
Specifically in the so-called rotor MOFs, these linkers can possess permanent quasi-free rotating dipole moments \cite{winston2008,bracco2017}, that could be arranged into arbitrary structures.

As a prototype model system for arrays of dipoles, we consider here a chain of equally spaced dipoles arranged along a helix. 
It has previously been shown, that the combination of long-range interactions and helical structures can lead to a variety of novel properties and dynamics \cite{schmelcher2011,zampetaki2013,zampetaki2015,zampetaki2015a,zampetaki2017,zampetaki2018,zampetaki2018a,siemens2020,siemens2021}, such as oscillating effective interaction potentials \cite{schmelcher2011}, band structure degeneracies \cite{zampetaki2015,zampetaki2015a}, or unusual transport properties in the presence of a driving field \cite{siemens2021}. 
Specifically dipoles in helical geometries have been studied in lattice models with long-range hopping \cite{wang2017}, and in classical setups with fixed dipole orientations \cite{pedersen2014}.

Motivated by the interesting effects found in the above works on model systems, we investigate here the configurations of a helical dipole chain with fixed positions of the dipoles and find the ground state (GS) to exhibit multiple crossovers between states that consist of a tunable number of superimposed helical dipole chains that wind around each other with either positive or negative chirality. 
Employing geometrical considerations that have previously been relevant in the field of Phyllotaxis \cite{adler1997}, we determine the underlying phase diagram and classify the resulting self-similar bifurcation diagram using fractions of the Farey sequence. 
The organizational principles of this emergent order and transitions between the occurring phases are explored.

\begin{figure}
\includegraphics[width=\columnwidth]{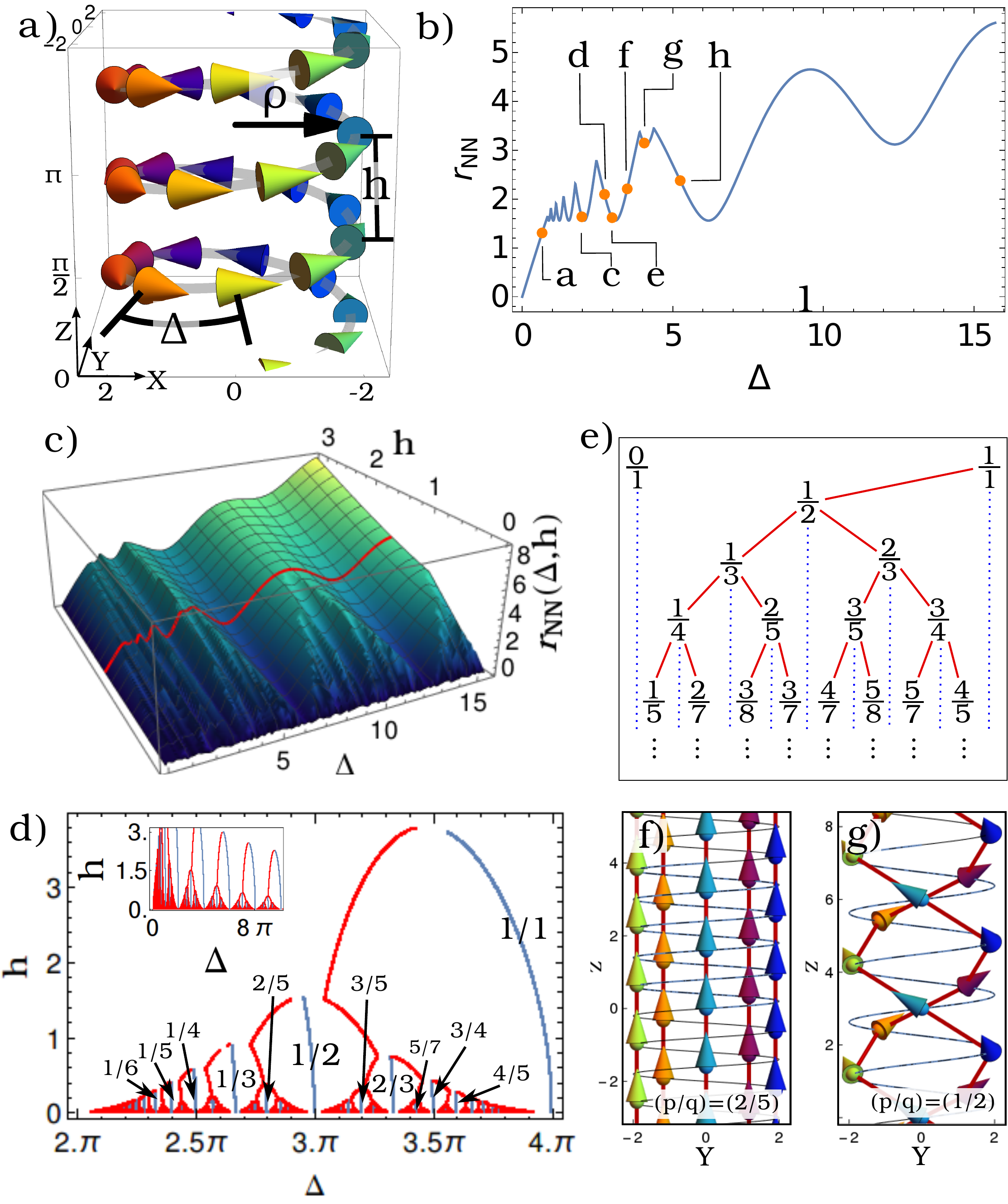}
\caption{\label{figure1r} (a) Visualization of the helical dipole chain and parameters for a helix radius $\rho=2$, a helix pitch $h=\pi/2$ and angular dipole-distance $\Delta=0.21\pi$. Coloring indicates the position within a winding. (b) Euclidean distance $r_{NN}$ to the nearest neighbor for $h=\pi/2$ and $\rho=2$ as a function of the dipole-distance $\Delta$. Configurations corresponding to the orange points are visualized in sub-figure \ref{figure1r} (a) and in sub-figures \ref{figure2r}(d-i). (c) Euclidean distance of nearest neighbors $r_{NN}$ as a function of the dipole-distance $\Delta$ and helix pitch $h$. The red line corresponds to configurations of Fig. \ref{figure1r}(b). (d) Bifurcation tree that shows the minima (blue) and maxima (red) of cross-sections of $r_{NN}(h,\Delta)$ for various $h$, corresponding to the valleys (blue) and ridges (red) of $r_{NN}(h,\Delta)$ in (c). The fractions $(p/q)$ classify the configurations between ridges. Note that the gaps in the drawing close to the bifurcation points reflect the subtle transition in the number of maxima (ridges) which is accompanied by intermediates of non-smooth derivatives. The inset depicts these extrema for a larger parameter regime. (e) Visualization of the Stern-Brocot tree. (f-g) Transition between a $(2/5)$ and a $(1/2)$ state by increasing the helix pitch $h$ from $0.8$ to $1.5$. Connections between NNs are indicated by red lines. }
\end{figure}

\paragraph*{Helical dipole chains}---
Our setup (see Fig. \ref{figure1r}(a)) consists of dipoles placed on a helix with radius $\rho$ and pitch $h$. 
The location of the $n$-th particle is then given by the following parametrization
\begin{equation}\label{eq:1}
\textbf{r}_n:=\left(
\begin{array}{c}
\rho\cos(n\,\Delta) \\
\rho\sin(n\,\Delta) \\
h\,n\,\Delta/2\pi
\end{array}
\right)
\end{equation}
where $\Delta$ is the angular distance between two dipoles along the helix. 
We consider an all-to-all dipole interaction. 
The interaction potential experienced by the $n$-th dipole is then given by 
\begin{equation}\label{Eq: potentialEnergy}
V_n = \sum_{\substack{i=-\infty \\i\neq n}}^{\infty} \dfrac{1}{4\pi}\left[\dfrac{\textbf{d}_i\,\textbf{d}_n}{r_{in}^3}-\dfrac{3(\textbf{d}_i\cdot\textbf{r}_{in})(\textbf{d}_n\cdot\textbf{r}_{in})}{r_{in}^5}\right]
\end{equation}
where $\textbf{d}_i$ is the dipole moment of the $i$-th dipole in the chain and $\textbf{r}_{in}=\textbf{r}_i-\textbf{r}_n$ is the separation vector between the dipoles $i$ and $n$ and $r_{in}$ is the corresponding magnitude. 
It should be noted, that due to the embedding of the dipoles in three-dimensional space, the nearest neighbor (NN) in Euclidean space does not necessarily agree with the corresponding next neighbor along the helical chain. 
As a result the alignment of the dipoles depends inherently on the geometrical parameters ($\rho$,$h$) of the helix and the chosen dipole angular spacing $\Delta$.

Our setup is scale invariant in the sense that for a given ratio of $\rho/h$, changes in the dipole strength $|\textbf{d}_i|$ or helix radius $\rho$ only scale the potential energy given by Eq. \ref{Eq: potentialEnergy} but do not lead to new equilibria. 
This allows us to normalize the helix radius $\rho$, as well as the dipole moments $\textbf{d}_i$. 
Without loss of generality, we therefore set $\rho=2$ and $|\textbf{d}_i|=1$. 
The relevant parameters describing our system are then $h$ and $\Delta$. 
If not explicitly stated otherwise, we focus on ground state configurations. 
These presented GS configurations of our many-body dipolar system are obtained as follows: 
First the GS is approximated by optimizing the energy with a simulated annealing method while constraining the dipole alignment to the surface of the cylinder spanned by the helix. 
Using the resulting configuration as an initial condition, the GS is found through optimization with a principal axis method.

\paragraph*{Phyllotaxis in cylindrical geometries}---
The considered system of equidistant particles on a 1D helix can also be interpreted as a cylindrical lattice where all lattice points can be accessed by a single generating helix. 
This set of cylindrical lattices has been studied in the past, and has been especially relevant in the field of Phyllotaxis \cite{adler1997} - the study of the arrangement of lateral organs in plants. 
In the context of Phyllotaxis, geometrical aspects of these lattices are used to explain the emergence of mathematical sequences, such as the Fibonacci sequence or the Lucas sequence in nature, e.g., in the arrangement of the scales of pine cones or pineapples. 
Patterns similar to those that emerge in phyllotactic systems can also be used to classify the ground state configurations in our helical dipole chain. 
We will now give a brief overview of the phyllotactic patterns emerging directly from the geometry of the setup.

\begin{figure}
\includegraphics[width=\columnwidth]{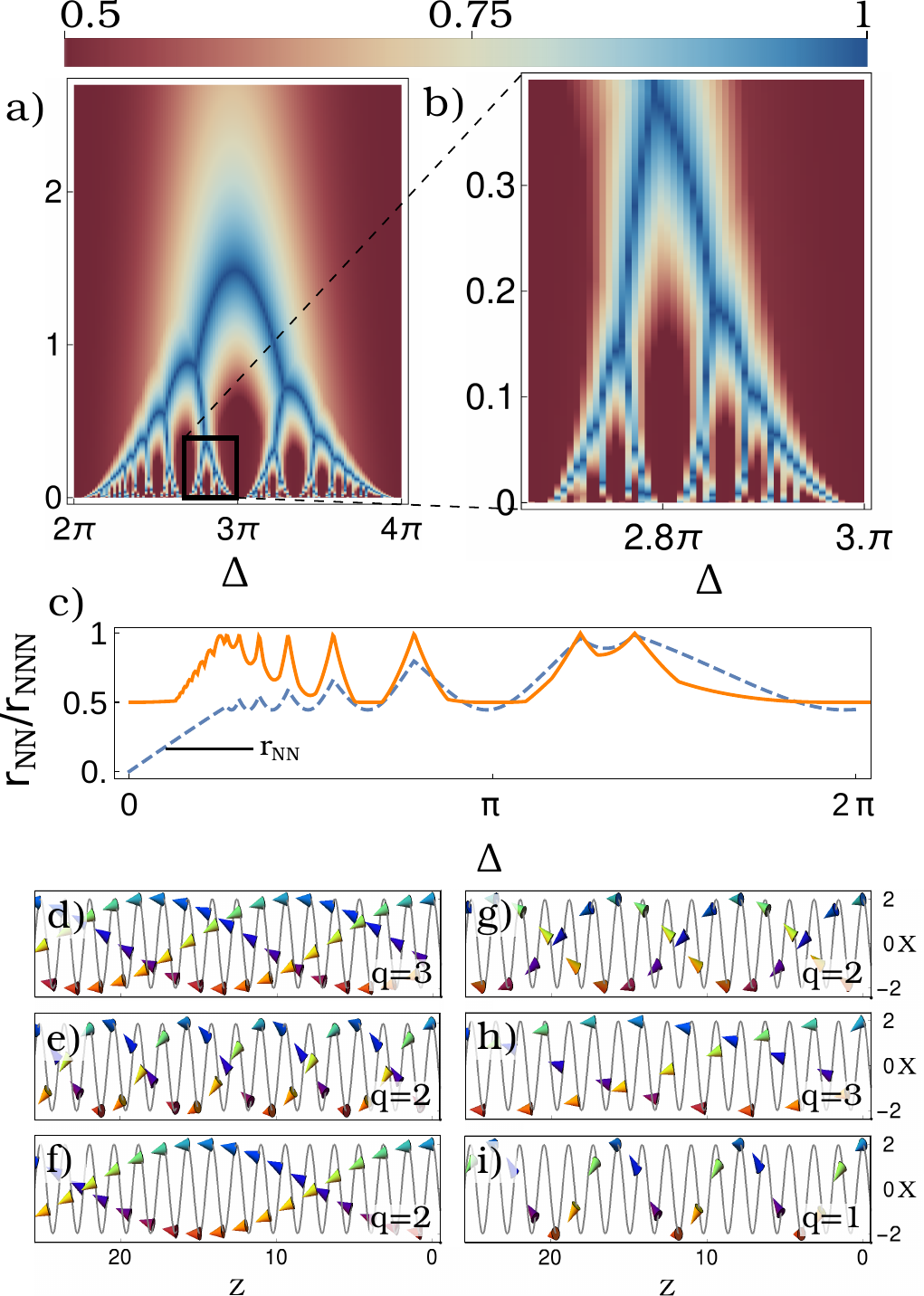}
\caption{\label{figure2r} (a) Ratio $(r_{NN}/r_{NNN})$ of the NN to the NNN distance as a function of $h$ and $\Delta$.  Interaction between different chains is (mostly) negligible in the red regions. (b) Zoom-in on (a). (c) Cross-section of $r_{NN}/r_{NNN}$ for $h=\pi/2$ along $\Delta$. The maxima of $r_{NN}/r_{NNN}$ coincide with the maxima of $r_{NN}$ (dashed blue line). The width of the peaks of $r_{NN}/r_{NNN}$ are proportional to $h\,\Delta$. (d-i) Side views of example configurations for $\Delta=0.63\pi, 0.87\pi, 0.95\pi, 1.11\pi, 1.29\pi, 1.67\pi$ respectively. The angular position of the dipoles within a helix winding, and thereby the chirality of the chains is encoded in the color. The parameter $q$ corresponds to the number of chains. }
\end{figure}

To understand these phyllotactic patterns, it is necessary to understand the so-called parastichy helices \cite{lee1998}. 
The parastichy helices are the secondary helices connecting all lattice points that can be reached by translation along the two shortest lattice vectors; either the NN vector $\textbf{r}_{NN}$ or the next-nearest neighbor (NNN) vector $\textbf{r}_{NNN}$. 
In nature, such as e.g. for the scales of pine cones, parastichy helices are usually much easier to visually identify than the underlying generating helix. 
Due to the lattice site indexing defined in Eq. \ref{eq:1}, the index $n$ of the lattice sites changes by a constant integer $s$ when translating along $\textbf{r}_{NNN}$ and by a constant integer $q$ when translating along $\textbf{r}_{NN}$. 
In Phyllotaxis (and the physical systems where similar geometrical considerations become important \cite{levitov1991,levitov1991a,nisoli2009a,nisoli2009}), it is these parastichy numbers $s$ and $q$ that are usually used to demonstrate mathematical sequences that govern the behavior of cylindrical lattices as a function of the parameters $h$ and $\Delta$. 
However, as we will show, in our setup the NNN interaction becomes negligible for large parameter regions (compare Fig. \ref{figure2r}(a-c)). 
Consequently, it can happen that $s$ changes, while our GS remains qualitatively unchanged when $h$ and $\Delta$ are varied. 
To uniquely classify the GS configurations of our helical dipole chains, we therefore need to deviate from the standard Phyllotaxis notation and classify the lattice configuration with the parameter $q$.

To understand how the NN index $q$ can describe arbitrary GS configurations of helical dipole chains, it is instructive to first focus on the case of $h=\pi/2$ and inspect the Euclidean distance $r_{NN}$ of NNs for varying $\Delta$ shown in Fig. \ref{figure1r}(b). 
When increasing $\Delta$ by starting at $\Delta=0$, $r_{NN}$ first increases almost linearly, and changes to an oscillatory behavior showing cusps at the maxima once $r_{NN}$ exceeds the helix pitch $h$. 
The cusp-like maxima of $r_{NN}(\Delta)$ correspond to sudden changes of the NN - and therefore to sudden changes of $q$. 
The overall behavior of $r_{NN}$ for arbitrary $h$ is similar to the above description for $h=\pi/2$. 
The NN distance as a function of $h$ and $\Delta$ i.e. $r_{NN}(\Delta,h)$ is shown in Fig. \ref{figure1r}(c). 
For reference, the intersection corresponding to Fig. \ref{figure1r}(b) is highlighted by a red line. 
For any cross-section with constant $h$ we observe, that once $r_{NN}$ exceeds $h$, the behavior changes from an almost linear increase to an oscillation with cusp-like maxima. 
In general, the number of extrema in each cross-section increases with decreasing $h$. 
Again, configurations with different parastichy number $q$ are separated by the maxima of the cross section. 
Consequently, configurations for different values of $q$ are separated by the ridges of $r_{NN}(\Delta,h)$. 
The positions of the ridges and valleys of $r_{NN}$ (i.e. positions of the minima and maxima of cross-sections of $r_{NN}(h,\Delta)$ for different $h$) are shown in the inset of Fig. \ref{figure1r}(d) for a broad range of values of $h$ and $\Delta$. 
We immediately recognize that their behavior follows a series of self-similar bifurcation trees. 
Each of the 'trees' is confined to a region of $2\pi m\leq \Delta \leq 2\pi (m+1)$ (for $m\in\mathbb{N}$) and the overall behavior is the same for all the trees. 
They characterize the same set of lattices - just with a different parametrization of the generating helix. 
In the $m$-th tree, there are $m-1$ empty windings without dipoles between next neighbor lattice sites along the helix (i.e. sites $n$ and $n+1$). 
On close examination, the trees differ in shape mainly by a scaling factor $1/\Delta$. 
This allows us to focus our analysis on a single tree. 
The structure of one such tree is shown Fig. \ref{figure1r}(d). 
Several features can be noticed here: 
For decreasing $h$, each ridge splits via a pitchfork bifurcation into a new valley and two ridges. 
When a new valley appears below a certain threshold value of $h$, it persists further for arbitrarily small values of $h$. 
In contrast, each ridge will only persist for some finite range of $h$ before separating into a valley and two new ridges.

The classification of the parameter regions between the extrema of $r_{NN}(h,\Delta)$ with integers $q$ follows a pattern. 
Firstly, when considering the $m$-th tree, for each (reduced) fraction $(p/q)\in[m,m+1]$ we can find a parameter region classified by a parastichy number $q$ that for $h\rightarrow0$ contracts towards $\Delta=2\pi(p/q)$ (shown in Fig. \ref{figure1r}(d)).  
Secondly, during a bifurcation of a ridge that separates a $(p_1/q_1)$ state from a $(p_2/q_2)$ state, the newly created state after the bifurcation can be described by the reduced fraction $(p_1+p_2)/(q_1+q_2)$. 
This is the same rule that generates new elements in the so-called Farey sequence \cite{boeyens2008}. 
And indeed, the possible configurations $(p/q)$ map exactly to the elements of the Farey sequence: 
with decreasing $h\Delta$ the Farey sequence is replicated to a higher order. 
For a better overview, the generic structure of the bifurcation tree is shown in Fig. \ref{figure1r}(e). 
The resulting tree is also known as the Stern-Brocot tree - a tree representation of the Farey sequence. 

An intuitive understanding for the corresponding lattice configurations can be gained by considering the $m$-th tree and the limit case of the circle ($h=0$). 
Placing dipoles on a circle with progressing angular winding of $\Delta=2\pi(m-1+p/q)$ will provide $q$ points on the simple circle $[0,2\pi]$ with a distance of $2\pi /q$. 
Therefore, for every rational number $(p/q)$ there exists a pitch $h_0$ so that for $h\leq h_0$ a helical GS configuration classified by the fraction $(p/q)$ can be found. 
The number of possible configurations $(p/q)$ decreases with increasing $h$. 
The reason, why certain $(p/q)$ states only persist for finite values of $h$ is illustrated in Fig. \ref{figure1r}(f-g). 
In both figures the angular positions of the dipoles are the same while $h$ is varied. 
The NN's are indicated by red connecting lines. 
Figure \ref{figure1r}(f) depicts a $(2/5)$ state for $h=0.8$. 
Above a certain value of $h$, the NN suddenly changes leading to the $(1/2)$ state shown in Fig. \ref{figure1r}(g) for $h=1.5$.

The Farey sequence, as well as geometric considerations similar to the ones above, have previously been employed in the description of physical systems, including layered superconductors \cite{levitov1991a}, repulsively interacting cylindrical lattices \cite{levitov1991}, and the domain wall dynamics in a magnetic cactus \cite{nisoli2009a,nisoli2009}. 
However, there are notable differences between the helical dipole chains and other Phyllotaxis related works, such as the above mentioned examples.  
In Phyllotaxis and related works, usually only those configurations with the closest packing density are of interest - corresponding to the ridges of $r_{NN}(h,\Delta)$. 
The configurations in between those ridges may (depending on the employed model) be accessible, but do not correspond to any equilibrium configuration.  
Consequently, in these works, it is the classification of these closest packing configurations which follows the Stern-Brocot tree and maps to the fractions of the Farey sequence.

Before continuing with the physics of interacting dipoles in helical geometries, a comment on the choice of coordinate system is in order.
Using the geometrical parameters $h$ and $\Delta$ allows to uniquely describe all possible cylindrical lattices of interest. 
In contrast, such a unique description of classifications is not achieved  with all coordinate systems. 
Using e.g. the ratio of primitive lattice vector magnitudes and the angle between those vectors is not sufficient, since in that case additional information relating the magnitude of one lattice vector to the circumference of the cylinder is required.

\paragraph*{Phyllotactic patterns in ground-state configurations of helical dipole chains}---
In this section, we will use the geometrical considerations and classification scheme provided above to describe the GS configurations of the helical dipole chains for arbitrary $h$ and $\Delta$. 
These GS configurations are for a large range of parameters dominated by NN interactions. 
This is demonstrated in Fig. \ref{figure2r}(a), which shows the ratio $r_{NN}/r_{NNN}$ of NN to NNN distances as a function of $h$ and $\Delta$. 
Comparing this to the bifurcation tree in Fig. \ref{figure1r}(d), we realize that the NNN interaction only becomes significant close to the maxima of $r_{NN}$. 
The cross section for $h=\pi/2$ in Fig. \ref{figure2r}(c) shows that the ratio $r_{NN}/r_{NNN}$ possesses pronounced peaks; in between those peaks flat regions emerge. 
In the flat regions, an asymptotic saturation tendency towards the value of $0.5$ can be observed (the relation $r_{NN}/r_{NNN}\geq0.5$ is guaranteed by the symmetric arrangement of dipoles within a single chain). 
As indicated by Fig. \ref{figure2r}(b), this behavior continues for arbitrary low $h$. 
This dominance of NN interactions allows us to directly translate the classification of lattice configurations with fractions of the Farey sequence to our helical dipole chain GS configurations whenever $r_{NN}\ll r_{NNN}$. 

Examples for various GS configurations in regimes of dominating NN interactions are shown in Fig. \ref{figure2r}(d-i). 
In these GS configurations, the dipoles generally align themselves with their NN's along several intertwined helical chains. 
Due to the symmetrical arrangement of dipole positions within a chain, dipoles will align in the plane spanned by the helix axis (z-axis) and the tangent vector $d\textbf{r}_n/d\Delta$. 
These intertwined helical dipole chains map exactly to the first parastichy helix. 
Therefore, the number of intertwined helical chains corresponds directly to the integer $q$ of the underlying lattice classification $p/q$. 
In addition to controlling the number of chains $q$, changing $h$ and $\Delta$ also allows to control the dipole density along the chain, as well as the length scale $\lambda$ (wavelength) on which the chains wind around each other. 
This change of $\lambda$ with varying $\Delta$ can be clearly seen in Figs. \ref{figure2r}(e-g). 
In Fig. \ref{figure2r}(f), close to the minimum of $r_{NN}$, NN dipoles show very gradual changes across the chain, thereby exposing the character of each separate chain. 
For smaller (larger) values of $r_{NN}$, the chains wind clockwise (counterclockwise) around each other (see Figs. \ref{figure2r}(e) and (g)). 
When $\Delta$ is increased further, such that $r_{NN}$ crosses a maximum, the GS configuration changes from counterclockwise chirality to a new set of chains with clockwise chirality.

\begin{figure}
\includegraphics[width=0.9\columnwidth]{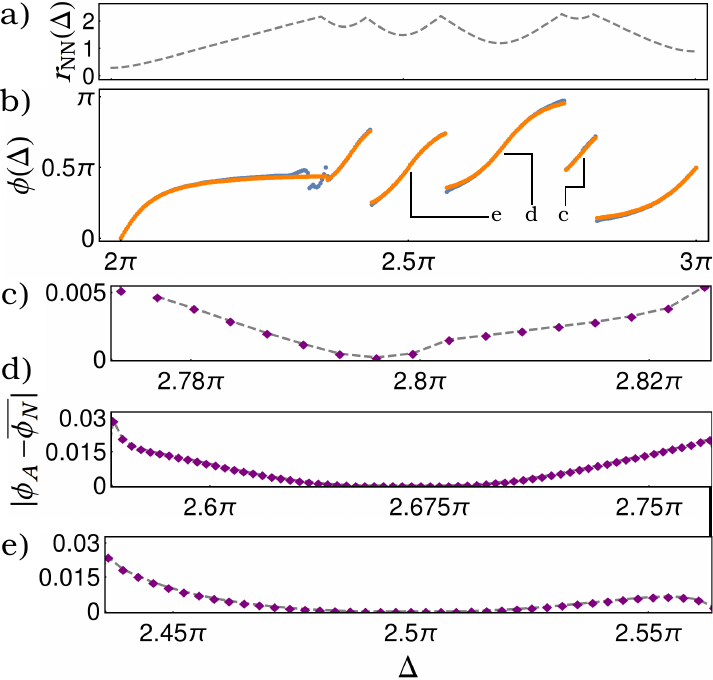}
\caption{\label{figure3r} (a) Distance to the nearest neighbor $r_{NN}(\Delta)$ for $\Delta\in[2\pi,3\pi]$ and $h=0.3$. (b) The analytically predicted angle $\phi$ (orange) compared to the numerically determined value (blue). Note that, to minimize edge defects, the numerical value corresponds to the average angle of dipoles from the bulk. Smooth regions are classified by the same fraction $(p/q)$, whereas large jumps in $\phi(\Delta)$ indicate a change to a configuration classified by a different fraction $(p/q)$. (c-e) The absolute difference $|\phi_A-\overline{\phi_N}|$ between the numerically and analytically determined angle for three of the regions with smoothly changing angle, i.e. three parameter ranges corresponding to parameter regions with different classification $(p/q)$. 
}
\end{figure}

The classification $(p/q)$ allows us to determine an analytical expression for the dipole alignments in a NN approximation. 
For a given state $(p/q)$ and a given helix geometry $h$ and $\Delta$, the angle $\phi$ between the dipoles and the helix axis (z-axis) is (approximately) given by the following equation: 
\begin{equation}
\label{eq:3}
\phi(\Delta,h)=-tan^{-1}\left[ \dfrac{hq\Delta}{2\pi \rho sin(q\Delta)}\right]\pm\dfrac{\pi}{2}
\end{equation}
where the term $\pm\pi/2$ selects an alignment parallel ($+$) or anti-parallel ($-$) to the helix axis. 
The accuracy of this approximation is demonstrated in Fig. \ref{figure3r}. 
As a representative example for the comparison shown in Fig. \ref{figure3r}, we consider a cross section through our parameter space with constant helix pitch $h$ and varying $\Delta$. 
The analytically approximated angles together with the corresponding angles obtained from numerical calculations are shown in Fig. \ref{figure3r}(b) for $h=0.3$ and $\Delta\in[2\pi,3\pi]$. 
Discontinuities (jumps) in the angle occur at the maxima of $r_{NN}$ (compare Fig. \ref{figure3r}(a)) when the configuration changes to a state with a different classification $(p/q)$. 
Within each of the regions where the angle changes smoothly the classification $(p/q)$ does not change. 
The difference between the two data-sets is for the most part very small. 
However, visible deviations consistently occur close to the maxima of $r_{NN}$. 
For a more detailed comparison of the deviations between the analytically and numerically determined angles, we show in Fig. \ref{figure3r}(c-e) their absolute difference for three of the `smooth' regions of Fig. \ref{figure3r}(b), i.e. three regions with different classifications $(p/q)$. 
In each of the three figures, the absolute difference between the analytically predicted and numerically calculated angles are shown. 
Close to the minima of $r_{NN}$ the analytical predictions are very accurate. 
With increasing distance from this minimum the error increases and reaches a maximum close to the maximum of $r_{NN}$. 
This is expected, since Eq. (3) is based on the fraction $(p/q)$ which is not well-defined for configurations in the immediate vicinity of the maxima of $r_{NN}$. 
This behavior can be summarized as follows: When the length scale on which the dipole chains wind around each other increases, the accuracy of the analytically predicted angles also increases.

\begin{figure}
\includegraphics[width=\columnwidth]{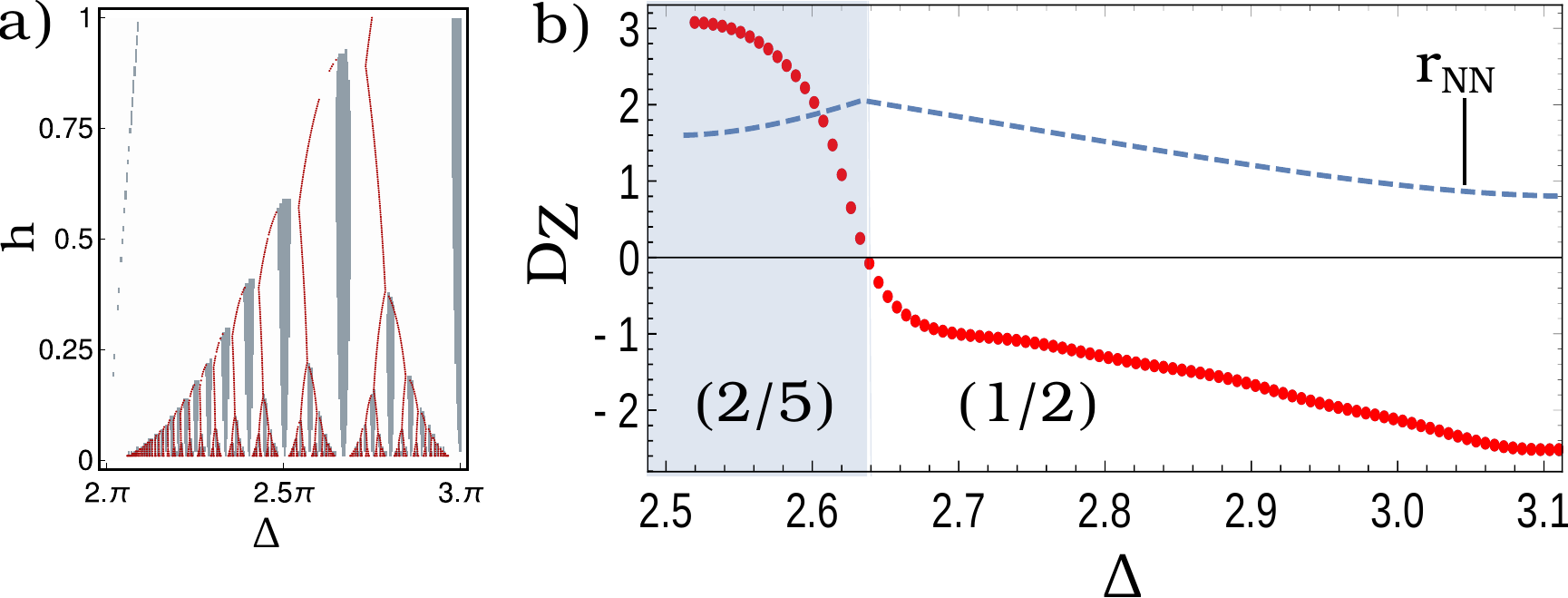}
\caption{\label{figure4r} (a) Classification of dipole alignments in the GS: In the white regions, a ferroelectric alignment is preferred, whereas anti-ferroelectric configurations are preferred in the gray regions, corresponding to configurations with very large wavelengths. For comparison, the bifurcation tree is shown in red. (e) Projected dipole density $D_Z$ along the helix axis (red) during a transition between a $(2/5)$ (blue area) and a $(1/2)$ (white area) state. For better identification of the transition state (i.e. maximum of $r_{NN}$), $r_{NN}$ is shown as a blue dotted line. }
\end{figure}

\paragraph*{Significance of interactions between chains}---
The interaction with NNN's can have significant effects on the dipole alignments in the GS - even in those parameter regimes where $r_{NNN}\ll r_{NN}$.
For parameter combinations where the NN interaction dominates, the NNN interaction still influences the relative alignment of the helical dipole chains to each other. 
They determine whether dipoles in neighboring chains are aligned parallel (ferroelectric (FE)) or antiparallel (anti-ferroelectric (AFE)) to each other. 
To study this, we compare the energies of FE and AFE states. 
As shown in Fig. \ref{figure4r}(a) for a specific parameter region, the AFE alignment is energetically favorable in the vicinity of $\Delta\approx2\pi(p/q)$ and sufficiently small $h$. For smaller wavelengths, the FE alignment becomes favorable. 
Note that the FE and AFE configurations, based on which the energies in Fig. \ref{figure4r}(a) were calculated, are obtained with the analytical formula given in Eq. (\ref{eq:3}) and not through numerical optimization.

Additionally, Fig. \ref{figure2r}(a) suggests that close to the ridges of $r_{NN}$ the NNN interaction becomes significant enough for the dipole configurations to deviate from the pure $(p/q)$ classification. 
To analyze this, we consider the transition between states for different values of $(p/q)$. 
As a representative example, we choose the transition from a $(2/5)$ to a $(1/2)$ state. 
It is achieved by varying $\Delta$ between $\Delta=2\pi(2/5)$ and $\Delta=2\pi(1/2)$. 
In our simulations, this was done by changing $\Delta$ in steps of $0.001$ and then relaxing the configuration with a Newton method. 
As a matter of fact, this transition leads to a drastic change from clockwise to counterclockwise chirality and vice versa. 
The transition could therefore result in a noticeable change of the dipole orientations. 
To analyze this, we introduce the projected dipole density $D_Z = \sum_i^N 2\pi\left(\textbf{d}_i\cdot \textbf{\^{e}}_Z\right)/hN\Delta$. 
For our example case, $D_z$ as a function of $\Delta$ is shown in Fig. \ref{figure4r}(b). 
Note, that since close to the transition state the FE alignment of neighboring chains is preferred we only consider FE configurations to study this transition.

In the course of the transition, $D_Z$ inverts its sign. 
Exactly at the maximum of $r_{NN}$, $D_Z$ is zero. 
The dipoles behave as follows: 
At $\Delta=2\pi(2/5)\approx2.513$ all dipoles are parallel to the helix axis. 
With increasing $\Delta$, the position of the NN changes and the angle between the dipole and the helix axis increases. 
Once $r_{NN}/r_{NNN}$ significantly deviates from $0.5$ and the NNN interaction becomes significant, the dipoles increasingly turn towards their NNN in the sense of a head-to-tail configuration. 
When $r_{NN}$ reaches a maximum, the dipoles are all aligned perpendicular to the helix axis. 
When $\Delta$ is increased further, the same behavior can be seen in reverse order until $\Delta=\pi$ is reached.

\paragraph*{Discussion and outlook}---
We have demonstrated that helical dipole chains exhibit a plethora of different equilibrium configurations yielding a variable number of tunable dipole chains winding around each other with either positive or negative chirality that are characterized by fractions of the Farey sequence. 
Varying the helical parameters yields a metamorphosis of these dipole states into each other. 
The observed chain formations, as well as the presence of FE and AFE GS configurations is consistent with previous studies \cite{rozenbaum1991,feldmann2008,brankov1987} of GSs of classical dipoles in 2D lattices.

For a large part of the parameter regimes the NNN interaction, and therefore the interaction between different chains, is negligible compared to the NN interaction. 
Within those regimes, an educated guess would be to expect that the mechanical and electrical response properties, as well as the information and energy transfer upon excitation is governed by the sum of the properties of these individual chains. 

It should be noted that our model system only exposes the dependence of dipole-chain properties on geometrical parameters. As such, all experimental realizations will most likely be affected by the presence of additional effects. 
A realization with the above mentioned MOFs will for example feature additional constraints \cite{gonzalez-nelson2019} on the dipole rotations: For one, certain rotation angles may be preferred due to the so-called torsion potential (an effective potential that can possess multiple minima as a function of the rotation angle). For another, the significant rotation of linkers in MOFs is typically only possible around one rotation axis while rotations around different axes are strongly constrained. 
Additional deviations from the studied setup could result from finite temperature and finite size effects. However, in our numerical calculations a significant deviation of dipole alignments from the bulk could only be observed for the first few dipoles of each chain.

%

\bibliographystyle{apsrev4-1}
\bibliography{txtest}

\end{document}